\documentstyle[12pt]{article}
\newcommand{\be}{\begin{equation}}
\newcommand{\ee}{\end{equation}}
\newcommand{\ba}{\begin{eqnarray}}
\newcommand{\ea}{\end{eqnarray}}

\date{}
\begin{document}
\begin{flushright}
Preprint SPbU-IP-95-09\\
May 1995\\
hep-th/9604145
\end{flushright}
\begin{center}
{\Large\bf Local realizations of $q$-Oscillators in Quantum Mechanics}\\
\vspace{1.cm}

{\bf A. A. Andrianov$^{a,b,}$\footnote{E-mail:  ANDRIANOV1@PHIM.NIIF.SPB.SU},
F. Cannata$^{c,}$\footnote{E-mail:  CANNATA@BO.INFN.IT},
J.-P. Dedonder$^{a,d,}$\footnote{E-mail: DEDONDER@PARIS7.JUSSIEU.FR}\\
and\\
M. V. Ioffe$^{b,}$\footnote{E-mail:  IOFFE@PHIM.NIIF.SPB.SU}}\\
\vspace{1.cm}

$^a$Laboratoire de Physique Nucl\'eaire, Universit\'e Paris 7-Denis Diderot\\
2 Place Jussieu, F-75251 Paris Cedex 05, France\\
\vspace{0.5cm}

$^b$Department of Theoretical Physics,Sankt-Petersburg State University,\\
 198904 Sankt-Petersburg, Russia,\\
\vspace{0.5cm}

$^c$ Dipartimento di Fisica and INFN, I-40126, Bologna, Italy \\
\vspace{0.5cm}

$^d$ Division de Physique Th\'eorique, IPN, F-91406 Orsay Cedex, France\\
\vspace{1.0cm}

{\large\bf Abstract}\\
\end{center}

\bigskip

{Representations of the quantum q-oscillator algebra are studied with
particular attention to local Hamiltonian representations of the
Schr\"odinger type. In contrast to the standard harmonic oscillators such
systems exhibit a continuous spectrum.
The general scheme of realization of the $q$-oscillator
algebra on the space of wave functions for a one-dimensional Schr\"odinger
Hamiltonian shows the existence of non-Fock irreducible representations
associated to the continuous part of the spectrum and directly related to
the deformation. An algorithm for the mapping of energy levels is described.}

\newpage
\vspace{1.cm}

\section{Introduction}
\hspace*{3ex} The $q$-deformed symmetries have been introduced
 \cite{Fad}, \cite{Dri} to characterize
the integrability of some lattice spin models (Yang-Baxter equations)
and conformal field theories \cite{KuRe}, \cite{Sk}.
Recently there have been
  many attempts to find examples of a $q$-deformed symmetry algebra in
 various quantum systems. The simplest algebra
which is important in quantum physics, the oscillator  Heisenberg-Weyl
algebra has been deformed in several ways \cite{Bied}, \cite{Kul} and its
representations have been formally classified \cite{Kul}, \cite{Rid},
\cite{Cha}.
The $q$-oscillators are often used as a tool to construct
more complicated quantum algebras by means of the Schwinger ansatz \cite{Kul},
\cite{Flor}. On the other hand, because of the significance
of the conventional oscillator in many areas of  modern quantum physics,
any new realization of the $q$-oscillator algebra
on the wave function space of a particular dynamical model is instructive
as to the role of this algebra for the physical systems.

It is our goal to describe different implementations of the
$q$-oscillator as a quantum system obeying the Schr\"odinger
equation. As well as for the harmonic oscillator ($q = 1$), the
connection between the values of different energy levels
and the corresponding wave functions \cite{Spir}
(or reflection and transmission coefficients \cite{ACDI}) is a consequence
of the $q$-oscillator algebra. One
realization was given in \cite{Spir} by means of the Darboux
transformation \cite{Darboux} supplemented with a dilatation of coordinates.
Since the Hamiltonians intertwined  by a Darboux transformation form
a super-Hamiltonian obeying the quantum mechanical superalgebra (in general,
a polynomial superalgebra, see  \cite{ACDI}, \cite{AIS} and references
therein) the additional dilatation leads to a $q$-deformation
of the supersymmetry itself. In order to reproduce a $q$-oscillator the
potentials in the Darboux connected Hamiltonians must
coincide up to a constant. This self-similarity property holds for the
conventional harmonic oscillator explaining its equidistant
energy spectrum. The dilated self-similarity condition \cite{Spir}, \cite{Shab}
naturally selects the potentials which yield the energy spectra
and wave functions of a $q$-deformed oscillator.

Here, we analyze the  possible realizations of
the $q$-oscillator on the space of wave functions for
a one-dimensional Schr\"odinger Hamiltonian. In our approach the local
Hamiltonian, in general, is not bilinear
 in creation and annihilation operators but rather belongs to the
universal enveloping $q$-oscillator algebra, i.e. to the algebra of
polynomials (analytic functions) of the generators. Thereby the $q$-oscillator
relations are considered as  a sort of $q$-deformed (nonlinear)
dynamical algebra.
In Sec. 2, we recall the algebra in the form \cite{Kul}, convenient
to build its representations in quantum mechanics.
The classification of representations is done in terms
of the central element and the second $q$-commutator of creation and
annihilation operators is derived. In Sec. 3, the decomposition of the
Hamiltonian realization introduced in
\cite{Spir} into irreducible $q$-oscillator
representations is developed and its polynomial generalization provided
by the $q$-deformation of a polynomial superalgebra is obtained.
Two types of
$q$-oscillator representations appear and, while the Fock representation
refers to the bound states, the non-Fock representations cover the continuous
spectrum.

In Sec. 4,  we formulate systematically a general
scheme of constructing a local
Hamiltonian of  Schr\"odinger type
with deformed spectrum generating algebras and the related  mapping of energy
levels is examined. The different forms of $q$-oscillator algebra are
described  and the constraints on their realization in terms of a
Schr\"odinger Hamiltonian are obtained.
In Sec. 5, we outline  possible applications and discuss
extensions of our approach onto other $q$-deformed algebras.

\section{$q$-oscillator algebra and its representations}
\hspace*{3ex} The deformation of a bosonic oscillator
can be defined in terms of
the $q$-commutator,\mbox{ $a a^+  - q^2 a^+ a$} where $a,\, a^+ = (a)^+$ are
annihilation and creation operators and $q$ is a real number which can
be taken positive
without loss of generality. It can be closed
in different forms among which we select  the following \cite{Kul},
\be
a a^+  - q^2 a^+ a = 1.    \label{com}
\ee
It is supplemented with the number operator $N$ such that,
\be
[N, a^+] = a^+; \quad [N, a] = - a  \label{num}
\ee
where the usual commutators are implied. For the harmonic oscillator
when $q = 1$, the ground state is a zero mode of the annihilation
operator, $a\psi_0 = 0$
 and the number operator $N$ can be normalized to be
$N = a^+ a$ so that the zero occupation number is assigned for
the ground state.

If $q \not= 1$ the number operator is no more bilinear
in $a, a^+$. However there is a central element\footnote{In fact,
it is not unique since any periodic function $\phi (N) =
\phi (N \pm 1)$ also belongs to the central element  subspace which
thereby consists
of any algebraic combinations of $\hat\zeta$ and $\phi(N)$. But for further
purposes it is sufficient to select out only one central element in the form
(\ref{char}).} given by \cite{Kul},
\be
\hat\zeta = q^{-2N} \left([N]_q - a^+ a \right), \label{char}
\ee
where the $q$-symbol $[N]_q$ is defined as
$$ [N]_q = \frac{1 - q^{2N}}{1 - q^2}.$$
The operator $\hat\zeta$ commutes with all generators of
the $q$-oscillator algebra as can be shown by the identities,
\be
a^+ F(N) = F(N-1) a^+;\quad a F(N) = F(N+1) a. \label{shift}
\ee
Therefore its eigenvalues $\zeta$ enumerate
the representations and, for a given representation
with a chosen $\zeta$, one can find the connection between the bilinear
operators $a^+ a, a a^+$  and the number operator $N$,
$$a^+a = [N]_q - \zeta q^{2N},$$
\be
a a^+ = [N +1]_q - \zeta q^{2N+2} \label{bil}
\ee
where the $q$-commutator (\ref{com}) has been used.
These operators commute with $N$ and their spectra
are generated by the spectrum of $N$. From Eqs.(\ref{bil})
the commutator can be evaluated,
\be
a a^+ - a^+ a
 =  q^{2N} \left( 1 - \frac{\zeta}{\zeta_c}\right),\label{sec}
\ee
where $\zeta_c$, the critical value of $\zeta$ that sets the commutator
to zero, is
$$\zeta_c = - \frac{1}{1 - q^2}.$$

Any representation can be described \cite{Kul,Rid}
in the basis of eigenfunctions of the
number operator with eigenvalues $\nu_n$, $N \psi_n = \nu_n \psi_n$.
As in the case of the conventional  harmonic oscillator,
due to relations (\ref{num}), all eigenstates can be built from one selected
state, $\psi_0 $ by means of the ladder operators,
$\psi_{n + 1} \simeq  a^+ \psi_n, \quad\psi_{n - 1} \simeq  a \psi_n$ and
hence
$\nu_{n \pm 1} = \nu_n \pm 1 $. For a chosen $\psi_0$
the nonequivalent representations are parametrized by the values of
$\nu_0$ running  within the unit interval,
$0 \leq \nu_0 < 1$ since the shift on
an integer number maps  one state to another one of the same representation.
We can redefine the number operator, $N = \tilde N + \nu_0 $ so that
the eigenvalues of $\tilde N$ become integer numbers. This redefinition is
compatible with the basic commutation relations (\ref{com}), (\ref{num})
and due to Eq.(\ref{char}) corresponds to the following change of the
central element,
\be
\hat\zeta' =\hat\zeta q^{2\nu_0}  - \zeta_c \left( q^{2\nu_0} - 1\right).
\label{zeta}
\ee
For $\zeta \not= \zeta_c$ any representation characterized by
two parameters $\nu_0,\zeta$ is equivalent to the representation labelled
by ${\nu'}_0 = 0, \zeta'$. In this case it is sufficient to
shift eigenvalues of the number operator
to integer numbers and to study eigenvalues of the
central element.  For $\zeta = \zeta_c$ the representations are
labelled by values of $\nu_0$.

Concerning the classification of $q$-oscillator representations in terms
of $\zeta$, there are three types of nonequivalent representations
\cite{Kul,Rid}: the
Fock representation $\zeta > \zeta_c$,
non-Fock representations for $\zeta < \zeta_c$
and the special representation for $\zeta = \zeta_c$.

The Fock representation is characterized by the existence of a ground state
of the number operator, zero mode of
the annihilation operator, $a \psi_0 = 0$. Since from Eqs.(\ref{com})
 and (\ref{sec}) and ${\nu'}_0 = 0$, we have
$\zeta' = 0$, the Fock representation is unique and can be built for
any $0 < q < \infty$.

The non-Fock representations may appear only if  $\,0 < q \leq 1$ and
thereby $\zeta_c < 0.$ The spectrum of
$N$ is unbounded from below. Consistency with Eqs.(\ref{bil})
requires $\zeta < \zeta_c$. Due to the relation (\ref{zeta}) there is a
one-parameter family of irreducible non-Fock representations.
Their parametrization can be realized either with the help of the
central element by fixing $\nu_0 =0$ or by means of the parameter
$0 \leq \nu_0 < 1$ for a fixed value of $\zeta$. In the following we make
use of the second part of the  alternative
and  for definiteness we set $\zeta = 2 \zeta_c$.

In the special representation (again for $0 < q \leq 1$) the creation and
annihilation operators commute (see Eq.(\ref{sec})) and it follows
from Eq.(\ref{com}) that the bilinear operators (\ref{bil})
become $c$-numbers,
\be
a a^+ = a^+ a = \frac{1}{1 - q^2} \equiv \rho^2. \label{spec}
\ee
This representation is generated by an unitary operator $U$ so that
$a = \rho U,\, a^+ = \rho U^+.$  The powers of  creation and annihilation
operators form a discrete subgroup of the $U(1)$ group. In this
representation the number
operator cannot be expressed as a function of $a, a^+$.

Thus any Hamiltonian realization of the $q$-oscillator algebra  can be
decomposed into the above irreducible representations.
In the following section we restrict ourselves to the $q$-oscillator
model with
a {\it local} (in $x$) Hamiltonian\footnote{There are also possibilities
to construct non-local Hamiltonians (see, for instance \cite{Bied},
\cite{Wess}) which we do not discuss in the present letter. } of the
Schr\"odinger type \cite{Spir}.

\section{Local Hamiltonian model of $q$-oscillator}
\hspace*{3ex} Let us construct the creation and annihilation operators
by means of Darboux operators (of first order in derivatives),
supplemented with the dilatation operator,
\be
a^+ = \left( - \partial_x + W(x)\right) T_q;\quad
a  = T_q^{-1}\left( \partial_x + W(x)\right) \label{dil}
\ee
where $W(x)$ is a real function and the  unitary dilatation
operator is defined by,
$$ T_q = \exp \left( \frac12 \eta  \{x, \partial_x\}\right),\quad
T_q \psi(x) = \sqrt{q} \psi(qx);\quad \eta \equiv \ln q .$$
Then the hermitian bilinear operators take the
Schr\"odinger Hamiltonian form,
\be
a^+ a = H_1 + k_1;\quad a a^+  = q^2 H_2 + k_2;
\quad H_{1,2} = - \partial_x^2 + V_{1,2} \label{mod}
\ee
where $k_i$ are constants. When $k_1 = k_2$  the Hamiltonians are
linked by the intertwining relations,
\be
H_1\, a^+ = q^2 a^+ H_2;\quad q^2 H_2\, a = a\, H_1. \label{inter}
\ee
The relation between potentials and the "superpotential" $W(x)$ are standard
for  Supersymmetrical Quantum Mechanics (SSQM) apart from the
dilatation contribution \cite{ACDI},
$$ V_1 (x) = W^2 (x) - W'(x) - k ;\quad
V_2 (x) = \frac{1}{q^2}W^2 ({x\over q}) +
\frac{1}{q}W' ({x\over q}) - \frac{k}{q^2}. $$
If we impose the conditions
$$V_1 = V_2, \quad k =  \frac{1}{1 - q^2}$$ the operators $a, a^+$  satisfy
the $q$-commutator relation (\ref{com}) which leads to the
$q$-self-similarity equation for $W(x)$:
\be
W'(x) + q W'(q x) + W^2(x) - q^2 W^2 (qx) =1 \label{W}
\ee
where the prime stands for the derivative with respect to $x$.
This equation can be considered on the entire axis, $x \in (-\infty ,
+\infty ),$ or on the semiaxis, $x \in [0 , \infty ).$

In the first case it has been shown \cite{Spir},
\cite{Shab} that for $0 < q \leq 1$
a nonsingular solution of this equation exists; its asymptotics,
$$ W(x)\vert_{\vert x\vert >> 1} = \pm\frac{1}{\sqrt{1 - q^2}} +
\mbox{\rm O}(\frac{1}{x^2})$$
leads to a decreasing potential,
$V(x) \sim 1/x^2,\quad \vert x\vert >> 1$.
The spectrum of the Hamiltonians of this type
must thus have a continuous part. If the solution $W(x)$
is chosen to take a positive constant
value for $x \rightarrow + \infty$  and a negative one for
$x \rightarrow - \infty$ it follows from Eqs.(\ref{dil}) that
the normalizable ground state $\psi_0 \sim \exp(- \int dx W(x))$ exists
and represents the zero-mode of the annihilation operator $a$:
in this case the $q$-oscillator model contains a bound state spectrum.

Let us decompose the $q$-oscillator representation given by the model
(\ref{dil}) - (\ref{W}) into
irreducible representations. According to the results of
Sec. 2 the bound state spectrum having the true ground state forms the Fock
representation, the continuous spectrum  consists of the set of non-Fock
representations. From the
$q$-oscillator relations (\ref{bil}) we find the number operator as
a function of the Hamiltonian for both cases,
\be
N = \frac{\ln \left[(1 - q^2)^2 H^2\right]}{4 \ln q}.
\ee
where the nonlinear operator relation can be interpreted in terms
of the spectral decomposition for the Hamiltonian $H$.
For the Fock representation this connection was found in \cite{Spir} and here
we extend it on the entire set of non-Fock representations for positive
energies.
Accordingly the central element can be defined as follows,
\be
\hat\zeta = \zeta_c (1 + \mbox{\rm sign}H),\quad H\psi = E \psi.
\ee
and in the Fock representation $\zeta = 0,\, E < 0$ and
$E_{n + 1} = q^2 E_n$ while in the non-Fock
representation one has $\zeta = 2\zeta_c,\, E > 0$. Both discrete and
continuous sequences of energy levels have $E = 0$ as the accumulation
point.

 The special representation could be realized on zero-energy states
(at the treshold between discrete and continuous spectra)
where $\zeta = \zeta_c$.
But for this particular model it can be proven
that the physical states\footnote{By physical states we understand
the wave functions which remain bounded at infinity.} for zero energy
do not exist because the two zero energy solutions have
increasing asymptotic behavior $\sim \sqrt{x}$.
Therefore the special representation
does not appear in the decomposition of the space of physical wave
functions.

We have thus shown that the $q$-oscillator
system with the local Hamiltonian (\ref{mod}) is composed
of two types of irreducible representations, in particular, the
continuous part of the spectrum is covered by non-Fock representations
parametrized by the second invariant within the interval $ 0\leq \nu_0 < 1$
corresponding to the energy interval
$ \vert\zeta_c\vert  \geq E > q^2 \vert\zeta_c\vert $,
\be
{\cal H} = {\cal H}_F
\bigoplus \int\limits_{0 \leq \nu_0 < 1}{} d\mu(\nu_0){\cal H}^{\nu_0}_{NF}.
\ee
where ${\cal H}$ denotes the appropriate Hilbert space of wave functions
(of both bound states and scattering states). The $q$-oscillator generators
act on scattering wave functions as pseudodifferential operators
in accordance with (\ref{dil}). It can be easily shown \cite{ACDI} that
their action preserves the scattering boundary conditions. A more
rigorous analysis will be done elsewhere.

We now present a novel generalization of the above $q$-oscillator model based
on the $q$-deformation of the polynomial supersymmetric algebra developed
in \cite{ACDI}. This algebra is generated by a Darboux differential operator
of higher order in derivatives dressed by a dilatation operator similarly
to (\ref{dil}). In order to
reproduce the $q$-oscillator algebra we are forced
to select the following algebras,
\be
a^+ a = H_1^n + \frac{1}{1 - q^{2n}};\quad
a a^+  = q^{2n} H_2^{n} + \frac{1}{1 - q^{2n}}. \label{polgen}
\ee
If $H_1 = H_2$ then
\be
a a^+ - q^{2n} a^+ a = 1.
\ee
The equations for $N$ and $\zeta$ are the same as for $n = 1$ provided
that one makes the substitution $q \longrightarrow q^n;\quad
H \longrightarrow H^n$. For $n \geq 2$ the polynomial in the
right hand side of (\ref{polgen}) has always
complex roots which corresponds to the primitive
Darboux transformations of second-order in derivatives
(for details see \cite{ACDI}).  There are two nonequivalent sets of
$q$-oscillator models corresponding to $n$ even or odd. The Hamiltonians
belonging to an odd algebra in general have the representation content
of the $n = 1$ model. In the even case it is clear that the Fock
representation is not involved into the decomposition of the related
$q$-oscillator wave function space since  $\Vert a^+ a\Vert >1/ (1 - q^{2n})$.

Let us consider, in particular, the case $n = 2$ where the creation and
annihilation operators read,
\be
a^+ = (a)^+ = \left( \partial^2_x - 2 f(x) \partial_x + b(x)\right)T_q.
\ee
From the intertwining relations (\ref{inter}) one finds,
\be
b(x) = f(x)^2 - f'(x)
 - \frac{f''(x)}{2f(x)} + \left(\frac{f'(x)}{2f(x)}\right)^2 +
\frac{1}{4f(x)^2 (1 - q^4)}.
\ee
The $q$-self-similarity equation now reads,
\ba
f(x)^2 + 2f'(x) + \frac{f''(x)}{2f(x)} - \left(\frac{f'(x)}{2f(x)}\right)^2 -
\frac{1}{4f(x)^2 (1 - q^4)} =\nonumber\\
q^2 f(qx)^2 - 2 q f'(qx) + \frac{f''(qx)}{2f(qx)} -
\left(\frac{f'(qx)}{2f(qx)}\right)^2 - \frac{q^2}{4f(qx)^2 (1 - q^4)},
\ea
This equation is essentially more complicated than (\ref{W})
and the existence of its regular solution has not been
yet established though for the semiaxis problem the linearization method
seems to be applicable and convergent.

Evidently from a regular nodeless solution $f(x)$ of
this equation one would
obtain the $q$-oscillator model with non-negative
Hamiltonian $ H = - \partial^2 + V(x)$ where
$$ V(x) = f(x)^2 - 2f'(x) +
\frac{f''(x)}{2f(x)} - \left(\frac{f'(x)}{2f(x)}\right)^2 -
\frac{1}{4f(x)^2 (1 - q^4)}.$$
Negative energy levels would only be allowed
for Hamiltonians unbounded from below because of the properties of non-Fock
representations. Such Hamiltonians would then be associated with singular
potentials.

Now we derive the systematic description
of deformed dynamical algebras which are realized on wave functions of
a Shr\"odinger operator and, in general, induce a {\it nonlinear} mapping of
energy levels.

\section{Generalized realizations of $q$-oscillator models}
\hspace*{3ex} To describe the $q$-oscillator algebra
in a generalized form we introduce new creation and annihilation operators
$A^+, A$ by means of a transformation preserving
the commutation relations with the
number operator (\ref{num}),
$$ a = F(N) A; \quad a^+ = A^+ F^*(N).$$
The function $F(x)$ can be chosen to be real since its phase factor
does not play any role in the relations to be derived below.
From (\ref{shift}) and (\ref{com})
the $q$-commutator with a new deformation parameter is reproduced if
\be
F^2 (N-1) = C_q F^2(N)
\ee
where $C_q$ is a positive $c$-number;
 this choice preserves the bosonic character of the algebra. The solution
of the previous equation reads,
$$ F^2(N) = C^{-N}_q \phi^{-1}(N);\quad \phi(N-1) = \phi(N).$$
The function $\phi$ is periodic, positive and in fact  takes
a definite value for a particular representation, thereby being an invariant.
From Sec. 2 it follows that $ \phi = \phi(\nu_0)$. The basic $q$-commutator
takes the form,
\be
A A^+ - q^2 C_q A^+ A = C_q^N \phi(N). \label{phi1}
\ee
The central element is modified as follows,
\be
\hat\zeta = q^{-2N} \left( [N]_q  - C_q^{1-N} \phi^{-1} (N) A^+ A \right).
\ee
The bilinear operators are modified as well,
\be
A^+ A = C_q^{N-1} \phi (N)\, a^+ a;\quad A A^+ = C_q^N \phi (N) \,a a^+,
\ee
as compared with Eqs.(\ref{bil}). The additional $q$-commutator  is built
with a new deformation parameter,
\be
A A^+ - C_q A^+ A = \left( 1 - \frac{\zeta}{\zeta_c}\right)
C_q^N q^{2N} \phi(N) \label{phi2}
\ee
The classification of representations is similar to that in Sec. 2
since we have not introduced any new algebra but have chosen different
elements of the same universal enveloping algebra as basic generators.
The special choice $C_q = 1/q$ leads to the algebra in \cite{Bied}
whereas $C_q = 1/q^2$ leads to the algebra introduced in \cite{Dam}.

Now we proceed to nonlinear
realizations of the spectrum generating deformed
algebras in  Quantum Mechanics.
Such realizations are of interest for the algebraic description
of physical systems which spectra are only approximately related to
the harmonic or $q$-harmonic oscillators.
We start from the generalized intertwining
relations,
\be
H_1 A^+ = A^+ g(H_2); \quad g(H_2) A = A H_1 \label{gen}
\ee
where $H_{1,2}$ are Hamiltonians of two quantum systems with related
energy spectra and wave functions,
$H_i\psi_i = E_i \psi_i;\quad E_1 = g(E_2), \quad \psi_1 = A^+ \psi_2$. In
the case of  polynomial SSQM we have $g(x) = q^2 x$.
In virtue of
(\ref{gen}), $\,[A^+ A, H_1] = [A A^+, g(H_2)] =0$.
If we assume in addition that the function g is invertible, we have
 $\, [A A^+, H_2] =0$; hence the bilinear operators commute with
the Hamiltonians and represent, in general, symmetry operators \cite{AIN}.
In  one-dimensional QM, they are functions of Hamiltonians,
\be
A^+ A = \sigma_1(H_1);\quad A A^+ = \sigma_2(H_2) \label{sig}
\ee
where $\sigma_i$ are arbitrary invertible functions.
It follows then that for analytic functions $f(z)$
\be
f(\sigma_1(H_1)) A^+ = \sum_{m=0}^{\infty} c_m
\sigma_1^m(H_1) A^+ = A^+ \sum_{m=0}^{\infty} c_m
\sigma_2^m(H_2) =
A^+ f(\sigma_2(H_2)) \label{sig2}
\ee
Let us now define the inverse functions,
\be
\sigma_i(\pi_i(z)) = \pi_i(\sigma_i(z)) = z,
\ee
and choose $f(z) = \pi_i(z)$. Then the
mapping function $g$ and its inverse function $g^-$
are determined by these functions,
\be
g(z) = \pi_1(\sigma_2(z)),\quad g^-(z) = \pi_2(\sigma_1(z)).
\ee
With the help of the inverse function one derives the second
set of intertwining
relations,
\be
 A^+ H_2 = g^-(H_1) A^+.
\ee
To find links between energy levels of the same dynamical
system  we impose the self-similarity condition, $H_1 = H_2$.
In this way we discover a deformed dynamical algebra of a Hamiltonian.
The ladder procedure connects different levels and
eigenstates belonging to the following sequence,
\ba
A^+:\quad
\cdots \longrightarrow  g^-(g^-(E)) \longrightarrow g^-(E)\longrightarrow E
\longrightarrow g(E)\longrightarrow g(g(E))\longrightarrow \cdots \nonumber\\
A:\quad
\cdots \longleftarrow  g^-(g^-(E)) \longleftarrow g^-(E)\longleftarrow E
\longleftarrow g(E)\longleftarrow g(g(E))\longleftarrow \cdots
\label{mapp}
\ea
For physical systems of oscillator type with Hamiltonians bound from
below this operators may generate the entire spectrum of a model whereas
on the continuous part of spectrum they connect only subsets of levels.

Imposing relations  (\ref{phi1}), (\ref{phi2}) on
the functions $\sigma_{1,2}$ given by (\ref{sig}),
we realize the $q$-oscillator algebra
provided that the following constraints for a representation
$[\zeta, \nu_0]$
 are fulfilled,\\
 for $C_q \not= 1$,
\be
\ln ( \sigma_2 (z) - C_q \sigma_1 (z)) = \left( 1 + \frac{2\ln q}{\ln
 C_q}\right)
\ln \left[ \sigma_2 (z) - q^2 C_q \sigma_1 (z)\right] -
\frac{2\ln q}{\ln C_q} \ln\phi(\nu_0) + \ln\left(1 - \frac{\zeta}{\zeta_c}
\right),
\label{sigm}
\ee
and  for $C_q = 1,\, \phi = 1$,
\be
\sigma_2 (z) - q^2  \sigma_1 (z) =1.
\ee
In the latter case the energy mapping reads,
\be
E \longrightarrow \pi_1\left( q^2 \sigma_1(E) + 1\right).
\ee
In particular, for $q=1$ this mapping represents the generalization
of the harmonic oscillator spectrum.

The content of irreducible
representations of the $q$-oscillator algebra
in a particular model depends on the position of
fixed points of the mapping (\ref{mapp}).
If a fixed-point energy value is higher than the ground
state energy the Fock representation exists and it is realized in between
them.
If there is no fixed points for finite energies then
the Fock representation spans the whole Hilbert space of wave
functions. On the
contrary if the fixed point coincides with the ground state energy the
non-Fock representations only appear in the decomposition into
irreducible representations.

\section{Conclusions and perspectives}
1.\quad The general strategy suggested by our approach is to find
the spectrum generating algebra
from the properties of the bound state spectrum
and scattering coefficients, i. e. to determine the functions
$\sigma_{1,2}$ satisfying the constraints (\ref{sigm}). In practice, it can
be done only approximately  and the required perturbation
theory will be studied
elsewhere. In order to find the related potential the inverse scattering
method is required to be extended onto potentials with $1/x^2$ asymptotics
at infinity \cite{Shab}.

\noindent
2.\quad There exist other generalizations \cite{Bon}
of the $q$-oscillator algebra of the following form
\be
a a^+ - \Phi_1^2 (N) a^+ a = \Phi_2^2 (N).
\ee
where $\Phi_i$ are supposed to be real functions sufficiently
regular.
In fact, one can redefine the basic elements of the universal enveloping
algebra,
\be
a = M(N) A ;\quad a^+ = A^+ M(N) ;\quad M^* = M;
\label{transf}
\ee
so to replace the function $\Phi_1(N)$ by a constant $c$. We restrict
ourselves to transformations which do not change the commutation
relations with the number operator.
The required function $M(N)$ obeys the following equation:
\be
\Phi_1(N) M(N+1) = c M(N) .
\ee
In a particular representation $\{ \zeta , \nu_0 \}$ its
solution is
\be
M(n + \nu_0) = M(\nu_0) c^{-n} \prod\limits_{l=0}^{n-1} \Phi_1 (l+\nu_0),
\ee
where $M(\nu_0)$ is an arbitrary function for $0 \leq \nu_0 < 1.$
Thereby one comes to the
$q$-deformed algebra (\ref{phi1}) but with the arbitrary function
$\Phi_2(N)$ on the righthand side in qualitative agreement
with \cite{Bon}. We stress however the additional freedom
to modify $\Phi_2(N)$ ,(\ref{transf}), which does not
change the enveloping algebra and was not  considered in
\cite{Bon}. Thus non-equivalent $q$-deformed
algebras are only those which cannot be related by
these "gauge" transformations.

\noindent
3.\quad We believe that the analysis of fixed points of functional mappings
(\ref{mapp}) can lead to a better understanding of the physical
signatures (see \cite{Bon} and references therein)
 of a $q$-deformed algebra. The
number of fixed points might represent a sort of topological invariant under
 "smooth"
perturbations of a potential. Still the problem of the implementation of
the special representation on wave functions of a local Hamiltonian
(a fixed point of energy mapping)
remains open.

\noindent
4. \quad
 We do not see any obstacles to extend our approach to the half-line
(radial) problem, at least, for the $S$-wave. In this
case one is allowed also to use solutions with singularities
at negative $x$.

\noindent
5. \quad
Finally we would like to mention the possibility to vary
the deformation parameter according to an external dynamics
(for example, a time-dependent $q = q(t)$ as an order
parameter describing the transition
from "confined", $q \geq 1$, to "deconfined", $q < 1$, phase).

\vspace{.5cm}
{\large\bf Acknowledgments}
\vspace{.5cm}

We thank Prof. E. V. Damaskinski for useful discussions and for the careful
reading of the manuscript. This work was supported by the University Paris
7-Denis Diderot   and by the RFBR grant No.96-01-00535 
and by the GRACENAS grant No.95-0-6.4-49.

After completing our paper we received
the paper \cite{Spir2} where a similar decomposition has been
described  without however introducing the notions of the central element
and $N$-operator.

\end{document}